\documentclass[aps,prl,groupedaddress,amsmath,amssymb,amsfonts,twocolumn,showpacs]{revtex4}
\usepackage{bm}
\usepackage{graphicx}
\usepackage{color}
\usepackage{psfig}

\begin{document}

\title{ Nonlinear magnetoconductance of a classical ballistic system}

\author{A. V. Andreev}

\affiliation{Department of Physics, University of Washington,
  Seattle, Washington, 98195, USA}

\author{L. I. Glazman}
\affiliation{W. I. Fine Theoretical Physics Institute, University of
Minnesota, Minneapolis, Minnesota 55455, USA}

\begin{abstract}
  We study nonlinear transport through a classical ballistic system
  accounting for the Coulomb interaction between electrons. The joint
  effect of the applied bias $V$ and magnetic field $H$ on the
  electron trajectories results in a component of the non-linear
  current $I(V,H)$ which lacks the $H\to -H$ symmetry: $\delta
  I=\alpha_{cl} V^2 H$. At zero temperature the magnitude of
  $\alpha_{cl}$ is of the same order as that arising from the quantum
  interference mechanism. At higher temperatures
  the classical mechanism is expected to dominate due to its
  relatively weak temperature dependence.
\end{abstract}


\maketitle

By Onsager symmetry~\cite{LandauLifshitz} the electric current
through a two terminal device in the linear regime is an even
function of the external magnetic field, $H$. The nonlinear current
does not necessarily satisfy this symmetry, as was recently observed
in experiments~\cite{Cobden,Marcus,Ensslin,Lindelof,Bouchiat}. In
fact it is violated already in the second order in the external bias
$V$: It was shown in Refs.~\onlinecite{Spivak2004,Sanches2004} that
the nonlinear current acquires an odd in $H$ component,
\begin{equation}\label{eq:V2H}
    \delta I = \alpha V^2 H.
\end{equation}
The violation of the $H\to -H$ symmetry in the nonlinear current is
associated with electron-electron interaction.

Electron-electron interactions result in inelastic scattering of
electrons.  The standard phase space argument shows that the
inelastic scattering cross-section varies as $\epsilon^2$, with
$\epsilon$ being the electron energy,  thus contributing to the
current only at order $V^3$ and higher. Therefore in order to
evaluate the coefficient $\alpha$ in Eq.~(\ref{eq:V2H}) one may
neglect inelastic processes and treat electrons as moving
independently in the presence of an effective potential which
depends on $V$ and $H$. Within such a
treatment~\cite{Spivak2004,Sanches2004} the bias induces an
additional electron density that is linear in the current and has an
odd in $H$ component. In the presence of interactions the additional
density changes the scattering potential, which results in the
nonlinear current (\ref{eq:V2H}).

In this approximation electron motion is  phase coherent. Therefore
a part of the nonlinear current is sensitive to the electron wave
interference. The corresponding contribution $\alpha_q$ to the
coefficient $\alpha$ in Eq.~(\ref{eq:V2H}) was studied in
Refs.~\onlinecite{Spivak2004,Sanches2004,Polianski,Deyo}. It is a
mesoscopic quantity whose $H$-dependence arises from the
interference pattern sensitivity to the magnetic flux threading
electron trajectories.

In addition to affecting electron interference the magnetic field
also bends electron trajectories. The corresponding contribution,
$\alpha_{cl}$, to the coefficient $\alpha$ is the subject of the
present work. In order to evaluate it electron motion can be treated
as classical.

We focus on the effect of the long range part of the Coulomb
interaction and show that its contribution to $\alpha_{cl}$ is
independent of the interaction parameter $e^2/\hbar v_F$ (here $e$
is the electron charge and the $v_F$ is the Fermi velocity). At
small $e^2/\hbar v_F$ this is the leading contribution to
$\alpha_{cl}$.

For an open two-dimensional ballistic dot the magnitude of the
classical contribution is
\begin{equation}\label{eq:alpha_cl_estimate}
    \alpha_{cl} \sim  \frac{e^4 n d^2}{\epsilon_F^2
    mc},
\end{equation}
where $d$ is the dot size, and $n$, $\epsilon_F$ and $m$ are the
electron density, Fermi energy and mass, respectively. In
Eq.~(\ref{eq:alpha_cl_estimate}) $e^4$ arises not from the
electron-electron interaction but from the coupling of electrons to
the magnetic field $H$ and potential $V$, and from the definition of
current $I$.

At zero temperature the interference contribution
is~\cite{Spivak2004} $ \alpha_q\sim \beta\frac{e^4 }{\nu E_T^2 \hbar
^2 c}$, where $\nu$ is the density of states at the Fermi level,
$\beta$ is the the dimensionless interaction constant and $E_T$ is
the Thouless energy. In the ballistic case, $E_T\sim \hbar v_F/d$,
and for $\beta \sim 1$ the coefficients $\alpha_{cl}$ of
Eq.~(\ref{eq:alpha_cl_estimate}) and $\alpha_q$ are of the same
order. Since $\alpha_q$ decays with temperature on the scale $T\sim
E_T$~\cite{Bouchiat} and $\alpha_{cl}$ is insensitive to $T$ for
$T<\epsilon_F$ we expect the classical contribution to
Eq.~(\ref{eq:V2H}) to dominate for $  T \gtrsim
E_T$~\cite{footnote}.

The classical and the quantum interference contributions can be
distinguished by their respective scales of the magnetic field
dependence. The linear $H$-dependence (\ref{eq:V2H}) holds as long
as the cyclotron radius significantly exceeds the system size $d$.
This limitation yields the magnetic field scale $H^*_{cl}\sim m v_F
c / (ed)$. The interference
contribution~\cite{Spivak2004,Sanches2004} is sensitive to the flux
threading a typical electron trajectory. Assuming that electron
motion is not chaotic we estimate the area of such a trajectory to
be $d^2$. Equating the flux piercing a typical trajectory to the
flux quantum $\Phi_0=hc/e$, we find the characteristic field for the
interference contribution is much smaller than the classical one,
$H^*_{Q}\sim \Phi_0/d^2\sim(\lambda_F/d)H^*_{cl}$, with $\lambda_F$
being the Fermi wavelength.

Similarly to the quantum interference
contribution~\cite{Spivak2004,Sanches2004} the magnitude and the
sign of the classical contribution depend on the sample geometry.

\begin{figure}[ptb]
\includegraphics[width=8.5cm]{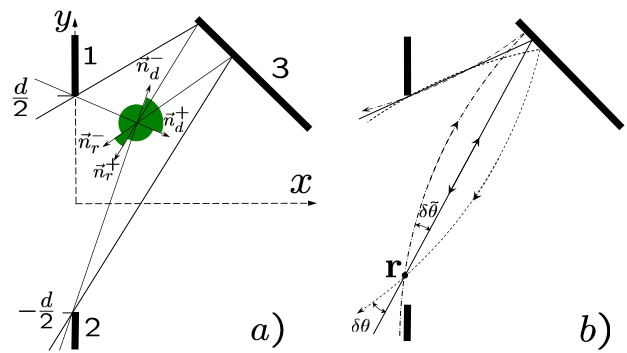}
\caption{Point contact, thick lines $1$ and $2$, with an adjacent
reflector, $3$. The reflector line is $y=L-x$ with $(L-d)/2\leq
x\leq (L+d)/2$. $(a)$: Separatrix trajectories and the
nonequilibrium electron distribution at $H=0$. Vectors
$\vec{n}^\pm_{d/r}$ represent velocity directions on the separatrix
trajectories. Angles between $\vec{n}^\pm_{d/r}$ and the $x$-axis
are denoted by $\theta^\pm_{d/r}$, respectively. $(b)$: Original
(dashed line) and modified (dash-dotted line) separatrix
trajectories at $H\neq 0$, and a separatrix at $H=0$ (solid line).}
\label{fig:setup}\label{figure}
\end{figure}

While the existence of the nonlinear current Eq.~(\ref{eq:V2H}) in a
classical system and the estimate, Eq.~(\ref{eq:alpha_cl_estimate}),
are quite general, below we concentrate on a specific setup
consisting of a ballistic point contact in a two-dimensional
electron gas (2DEG) with an adjacent ``reflector", see
Fig.~\ref{fig:setup}. The reflector creates the spatial asymmetry
necessary for ``rectification" current $\sim V^2$. (The linear
electron transport in systems of this type was studied
experimentally in Ref.~\onlinecite{Westervelt}.) Electron motion in
such a system is non-chaotic. This is essential for the validity of
our results and the estimate (\ref{eq:alpha_cl_estimate}). The case
of chaotic classical motion is beyond the scope of this work.

In the classical description of electron transport the local
momentum distribution function of electrons,
$n(\mathbf{p},\mathbf{r})$, plays a central role. The current $I$
across the contact is
\begin{equation}\label{eq:current_n_short}
I=\int d{\bf S} \cdot \mathbf{j}(\mathbf{r})\, , \quad
\mathbf{j}(\mathbf{r})=\frac{e}{m}\int \frac{d^2 p}{(2\pi
\hbar)^2}\, \mathbf{p}\, n(\mathbf{r},\mathbf{p}),
\end{equation}
where $\mathbf{j}(\mathbf{r})$ is the current density and the vector
$d {\bf S}$ is normal to the contact line. A voltage bias applied to
the contact results in a non-equilibrium, anisotropic electron
distribution $n({\bf r}, {\bf p})$ yielding a finite current $I$.

In order to determine the nonlinear $I$-$V$ characteristic, we need
to find the steady-state nonequilibrium distribution function
$n({\bf r}, {\bf p})$.  This can be done by solving the classical
equations of motion for electrons moving in the presence of a
self-consistent electrostatic potential $\phi(\mathbf{r})$. The
initial conditions for the electron dynamics correspond to the
equilibrium distributions deep inside the leads with electrochemical
potentials differing by $eV$. Conservation of electron energy
enables us to express the electron distribution at any point ${\bf
r}$ in the form
\begin{equation}\label{eq:n_p_LR_short}
n({\bf r},{\bf p})=\left\{\begin{array}{cc}
  f(\epsilon_p-\mu_0+e\phi({\bf r})-eV), & {\bf p}\in L({\bf r}), \\
  f(\epsilon_p-\mu_0+e\phi({\bf r})), & {\bf p}\in R({\bf r}). \\
\end{array} \right.
\end{equation}
Here $\epsilon_p=p^2/(2m)$ is the electron kinetic energy, $\mu_0$
is the equilibrium chemical potential,
$f(\epsilon)=[\exp(\epsilon/T)+1]^{-1}$ is the Fermi function, and
we assumed that the voltage $eV$ is applied to the left lead. The
solution of the electron equations of motion is encoded in the
shapes of the complementary  momentum space domains, $L({\bf r})$
and $R({\bf r})$. Electrons with momenta in these domains arrive to
${\bf r}$ from the left and right lead, respectively.

The electric potential $\phi({\bf r})$ is determined from the
Poisson equation. Its solution is greatly simplified if the
screening radius is short compared with the geometrical
characteristics of the setup. Under that condition, the charge
density is unchanged by the applied bias. A further simplification
is possible if the width $w$ of the depletion layers confining the
2DEG is small compared to the geometrical features of the system.
The electron density increases from zero in the depletion region to
its bulk value $n_0$ over a distance of order $w$, see
Ref.~\onlinecite{Glazman_Larkin}. Thus at small $w$ we may assume a
constant electron density,
\begin{equation}\label{eq:neutrality_new}
    \int\frac{d^2 p}{(2\pi \hbar)^2}\,n(\mathbf{p},\mathbf{r})=n_0,
\end{equation}
inside the 2DEG. This is a reasonable approximation for devices
fabricated using the local oxidation
method~\cite{Ensslin_fabrication}. The latter enables fabrication of
structures with very small lateral depletion widths, $w\sim 150 \AA
\sim \lambda_F$~\cite{Ensslin_fabrication}.

In order to find the current with the accuracy $\propto V^2$, we
will solve the transport problem defined by
Eqs.~(\ref{eq:current_n_short})--(\ref{eq:neutrality_new}) using the
following iterative procedure. At the first step, we find the
electron trajectories at $\phi ({\bf r})=0$. This defines the zeroth
iteration for the momentum domains, $L^{(0)}({\bf r})$ and
$R^{(0)}({\bf r})$. The distribution function in this approximation
depends on $\phi({\bf r})$ only explicitly, through the arguments of
the equilibrium distribution functions in
Eq.~(\ref{eq:n_p_LR_short}). Substituting $n({\bf p},{\bf r})$ into
Eq.~(\ref{eq:neutrality_new}) we obtain an equation for
$\phi^{(1)}({\bf r})$ which is valid to the first order in $V$.
After finding $\phi^{(1)}({\bf r})$, we determine corrections to the
electron trajectories and the corresponding corrections to the
momentum domains. The corrected momentum domains $L({\bf r})$ and
$R({\bf r})$ and $\phi^{(1)}({\bf r})$ determine the distribution
function via Eq.~(\ref{eq:n_p_LR_short}). Substitution of the latter
into Eq.~(\ref{eq:current_n_short}) gives the current to the second
order in $V$.

For our device geometry each of the domains $L({\bf r})$ and $R({\bf
r})$ consists of two separate sectors. The two sectors in $L({\bf
r})$ correspond to trajectories arriving to ${\bf r}$ from the left
lead either directly through the contact, or upon reflection from
the obstacle, see Fig.~\ref{fig:setup}. The boundaries of domains
$L({\bf r})$ and $R({\bf r})$ are defined by a set of four
energy-dependent angles, $\theta^\pm_{d/r}(\mathbf{r},\epsilon_p)$
between the $x$-axis and the velocity vector at point ${\bf r}$ for
the separatrix trajectories. These trajectories pass through the
edge points of the contact and through point ${\bf r}$. The indices
$^\pm_{d/r}$ denote the trajectories that pass through the
top($^+$)/bottom($^-$) edge point of the contact and arrive at ${\bf
r}$ either directly ($_d$) or upon reflection off the obstacle ($
_r$).

To evaluate the current $I$ via Eq.~(\ref{eq:current_n_short}), we
express the $x$-component of the current density, $j_x({\bf r})$, at
the contact line, $x=0$, in terms of the angles
$\theta_{d/r}^{\pm}({\bf r},\epsilon)$:
\begin{subequations}\label{eq:current barrier_short}
\begin{eqnarray}
j_x({\bf r})&=&j_x^d({\bf r})+j_x^r({\bf r})\,,
\label{jdjr}\\
j_x^d({\bf r})&=&\frac{e\nu_0}{\pi}\int_{\epsilon_-}^{\epsilon_+}d
\epsilon \, \sqrt{\frac{2 \epsilon}{m}} \,
    \sin\theta \Big|_{\theta^+_d({\bf r},\epsilon)}^{\theta^-_d({\bf
    r},\epsilon)}\,,\label{jd}\\
j_x^r({\bf r})&=&\frac{e\nu_0}{\pi}\int_{\epsilon_-}^{\epsilon_+}d
    \epsilon \, \sqrt{\frac{2 \epsilon}{m}}
    \sin\theta\Big|_{\theta^-_r({\bf r},\epsilon)}^{\theta^+_r({\bf
    r},\epsilon)}\,,\label{jr}
\end{eqnarray}
\end{subequations}
and treat $I$ as the difference:
\begin{equation}
I=I_d-I_r\,,\quad I_d=\int dS j_x^d({\bf r})\,,\quad I_r=-\int dS
j_x^r({\bf r}). \label{Itotal}
\end{equation}
Here $\nu_0=\frac{m}{2\pi \hbar^2}$ is the electron density of
states per spin projection, $\epsilon_-=\mu_0-e\phi({\bf r})$,
$\epsilon_+= \mu_0-e\phi({\bf r})+eV$, and the integrals over $dS$
go along the contact line. We set $T=0$ for brevity. The
generalization to finite temperature is straightforward; the scale
for the temperature dependence of the current is set by the Fermi
energy, $\mu_0$.

Initiating the iterations, we find the angles
$\theta^{\pm}_{d/r}=\theta^{\pm, (0)}_{d/r}(\mathbf{r}, \epsilon)$
at $V=0$. It is clear from Eq.~(\ref{eq:current barrier_short}) that
to find the linear term in the $I$-$V$ characteristic, it is
sufficient to know $\theta^{\pm, (0)}_{d/r}(\mathbf{r}, \epsilon)$
only at the Fermi energy. Equation (\ref{eq:current barrier_short})
also indicates that the spatial dependence $\phi({\bf r})$ does not
affect the linear conductance~\cite{Shekhter}.

Next, using Eqs.~(\ref{eq:n_p_LR_short}) and
(\ref{eq:neutrality_new}) we obtain the first iteration for the
potential $\phi({\bf r})$,
\begin{equation}\label{eq:phi_theta_short}
    \frac{\phi^{(1)}({\bf r})}{V}=1-
    \frac{\theta^{+,(0)}_d({\bf r}) - \theta^{-,(0)}_d({\bf r})
 -\theta^{+,(0)}_r({\bf r}) +\theta^{-,(0)}_r({\bf r})}{2\pi}.
\end{equation}
Here the angles $\theta$ are evaluated at the Fermi level,
$\epsilon=\mu_0$. This simplification is possible because the
electron distributions in the domains $L({\bf r})$ and $R({\bf r})$
differ from each other only within a narrow energy strip of width $
eV$ around the Fermi energy.

Finding the electric field from Eq.~(\ref{eq:phi_theta_short}) we
determine the correction to electron trajectories in the linear
order in $V$. This defines the first iteration of the angles,
$\theta^{\pm,(1)}_{d/r}({\bf r},\epsilon)$. To find the current to
the order $V^2$ we substitute
\begin{equation}\label{eq:theta_iteration}
    \theta^{\pm}_{d/r}({\bf r},\epsilon)=\theta^{\pm,(0)}_{d/r}({\bf
r},\epsilon)+\theta^{\pm,(1)}_{d/r}({\bf r},\mu_0)
\end{equation}
into Eqs.~(\ref{eq:current barrier_short}) and (\ref{Itotal}). Note
that we set $\epsilon \to \mu_0$ in $\theta^{(1)}$ because
$\theta^{(1)}$ is already proportional to $V$.

In the absence of magnetic field and at zero bias the electron
trajectories are straight lines. Therefore the angles $\theta^{(0)}$
do not depend on the energy and can be easily found from the
geometric construction, see Fig.~\ref{fig:setup}:
\begin{subequations}
\label{eq:angles_zero}
\begin{eqnarray}\label{eq:angles_zero_a}
    \theta^{\pm,(0)}_d({\bf r})&=&\mathrm{Im} \ln[ x+iy \mp id/2], \\
    \label{eq:angles_zero_b}
     \quad \theta^{\pm,(0)}_r({\bf r})
  &=&\mathrm{Im} \ln[ x+iy-L(1+i) \pm d/2].
\end{eqnarray}
\end{subequations}

The first iteration $\theta^{(1)}(r)$ is determined from the condition that the
velocity direction at point ${\bf r}$ must change in the presence of the external
force $\nabla\phi^{(1)}$ in such a way that the edge point of the contact still
belongs to the separatrix trajectory. A perturbative solution of Newton's
equations of motion gives,
\begin{equation}\label{eq:delta_theta_answer}
    \theta^{(1)}=\frac{1}{2 l_0 v_F^2} \frac{e}{m} \int^{l_0}_0 dl \int^{l}_0
    dl' \nabla_\perp \phi^{(1)}(l'),
\end{equation}
where $l_0$ is the length of the unperturbed trajectory, $l$ is the
coordinate along the trajectory, and $\nabla_\perp \phi^{(1)}(l')$
is the component of the electric potential gradient that is
perpendicular to the trajectory. Substitution of the angles
Eq.~(\ref{eq:angles_zero}) found in the zeroth order into
Eq.~(\ref{eq:phi_theta_short}) and subsequent solution of
Eq.~(\ref{eq:delta_theta_answer}) yield corrections
$\theta_{d/r}^{\pm,(1)}$ which are proportional to $V$ and depend on
the geometry of the device. Substitution of the found
$\theta_{d/r}^{\pm,(0)}$ and $\theta_{d/r}^{\pm,(1)}$ into
Eq.~(\ref{eq:current barrier_short}) yields the $I$-$V$
characteristic with linear and $\propto V^2$ terms:
\begin{equation}\label{eq:total_current_short}
    I =G_{Sh}V \left[ 1 -\frac{d}{4\sqrt{2}L}\left(1 +\frac{1}{\pi \sqrt{2}}
    \frac{eV}{\mu_0} \right) \right],
\end{equation}
where $G_{Sh}=2e^2 d/(\pi \lambda_F)$ is the Sharvin conductance of
the point contact~\cite{Sharvin}. The rectification ($\propto V^2$)
term and the corrections to the linear Sharvin conductance exist
only due to the electron trajectories which are reflected by the
obstacle back into the contact. The relative weight of such
trajectories, contributing to Eq.~(\ref{jr}), is $\propto d/L$. In
writing Eq.~(\ref{eq:total_current_short}) we took the limit $d\ll
L$, in order to simplify the result. In this limit the rectification
term arises solely from the energy dependence of the velocity
$\sqrt{2\epsilon/m}$ in the integrand of Eq.~(\ref{eq:current barrier_short}).

Now we evaluate the influence of the magnetic field, assumed to be perpendicular
to the plane of motion. In this case it is convenient to evaluate the
``back-current'' $I_r$ in Eq.~(\ref{Itotal}) in a different way. Equations
(\ref{jr}) and (\ref{eq:total_current_short}) express the ``back-current'' in
terms of the directions of velocities of electrons that are already reflected
from the obstacle and head towards the contact. We may, instead, evaluate the
back-current by accounting for those trajectories of electrons that head towards
the obstacle and will subsequently scatter back in the contact. We introduce the
angles $\tilde{\theta}^\pm_r({\bf r})$ between the $x$-axis and the velocity
direction at point ${\bf r}$ for the modified separatrix trajectories that
\textit{start} from ${\bf r}$ and, upon reflection from the barrier,
\textit{arrive} to the top/bottom edge point of the contact, see
Fig.~\ref{fig:setup} b). In terms of these angles, the back-current is
\begin{equation}
I_r^\prime=\frac{e\nu_0}{2\pi}\int dS\int_{\epsilon_-}^{\epsilon_+}d \epsilon \,
\sqrt{\frac{2 \epsilon}{m}} \sin\theta\Big|_{\tilde{\theta}^-_r({\bf
r},\epsilon)}^{\tilde{\theta}^+_r({\bf r},\epsilon)}\,. \label{Irprime}
\end{equation}
Here the integration is performed along the contact line. The
stationary-state current conservation law dictates
$I_r^\prime=I_r$. It is convenient for us to replace $I_r$ in
Eq.~(\ref{Itotal}) by $(I_r+I_r^\prime)/2$. Upon this substitution, we find:
\begin{equation}\label{eq:current barrier_short_effective}
 I=\frac{e\nu_0}{2\pi}\int \!dS\!\int_{\epsilon_-}^{\epsilon_+}\!\!\!d \epsilon \,
 \sqrt{\frac{2 \epsilon}{m}} \,
    \left[ 4-
    \sin\theta\Big|_{\theta^+_r({\bf r},\epsilon)}^{\theta^-_r({\bf
    r},\epsilon)}\!\!\!-
    \sin\theta
    \Big|_{\tilde{\theta}^-_r({\bf r},\epsilon)}^{\tilde{\theta}^+_r({\bf r},\epsilon)}
    \right]\,.
\end{equation}
To arrive at this equation we used the fact that on the contact line
$\theta^{\pm}_d({\bf r}) =\pm \pi/2 + O(H, V)$. This enabled us to
set $\sin\theta^\pm_d({\bf r},\epsilon) =\mp 1$ in Eq.~(\ref{jd}).

The modified separatrix trajectories  coincide with the
time-reversed original separatrix trajectories in the opposite
magnetic field and unchanged electric potential. Therefore the
modified angles can be expressed in terms of the original ones as
$\tilde{\theta}^\pm_r({\bf r}, H)=\theta^\pm_r({\bf r}, -H)-\pi $.
Thus Eq.~(\ref{eq:current barrier_short_effective}) explicitly shows
the $H\to -H$ symmetry of the {\sl linear} current, in agreement
with the Onsager relations.

In the presence of a magnetic field, the trajectories become curved
even at zero bias.  We denote the deviation of the angle $\theta$
from its value at $H=0$ by $\delta \theta$. For a trajectory
originating from ${\bf r}_i$ and arriving at ${\bf r}_f$ without
reflection off the barrier in a weak magnetic field we have
\begin{equation}\label{eq:delta_theta_H_short}
    \delta \theta \approx |{\bf r}_f-{\bf r}_i|/(2R_c),
\end{equation}
where $R_c=mv c/(eH)$ is the cyclotron radius. The angles
$\theta^{\pm(0)}_{d}(z)$ acquire the correction $\delta\theta^{\pm(0)}_{d}({\bf
r})=|x+iy \mp id/2|/(2R_c)$. Though the expressions for the corresponding
corrections to the reflected angles $\delta \theta^{\pm,(0)}_r$ are more
cumbersome they can be straightforwardly obtained from geometric considerations
using Eq.~(\ref{eq:delta_theta_H_short}). Substituting these corrections into
Eq.~(\ref{eq:phi_theta_short}) we obtain the correction to the induced potential
due to the magnetic field. It has an especially simple form on the contact line,
$x=0$,
\begin{equation}\label{eq:delta_phi_H_contact_short}
    \delta \phi_H^{(1)} (x=0,y)=\frac{V}{4\pi
    R_c}\left[2 y+\frac{3d}{2\sqrt{2}}\right].
\end{equation}
Although the second term here is caused by the reflected
trajectories, the length $L$ defining  the barrier position drops
out for $L \gg d$. Indeed, from Eq.~(\ref{eq:delta_theta_H_short})
it follows that the corrections to the reflected angles are $\delta
\theta^\pm_r \sim L/R_c$, whereas their difference entering the
potential via Eq.~(\ref{eq:phi_theta_short}) has an additional
smallness of $d/L$.

Next we evaluate the current across the contact using Eq.~(\ref{eq:current
barrier_short_effective}). It depends on magnetic field due to the $H$-dependence
of the integration limits, $\epsilon_\pm$, as described by
Eq.~(\ref{eq:delta_phi_H_contact_short}), and of the angles $\theta$,
$\tilde\theta$ entering in the integrand. The finite-$H$ corrections to the
angles $\theta$ and $\tilde{\theta}$ arise from two effects: i) the direct
bending of electron trajectories by the Lorentz force, and ii) the $H$-dependence
of the electric potential, Eq.~(\ref{eq:delta_phi_H_contact_short}).  For our
geometry these effects are small: The corrections to $\theta$ and
$\tilde{\theta}$ due to the first effect are opposites of each other, as
illustrated in Fig.~\ref{fig:setup} b), and drop out from Eq.~(\ref{eq:current
barrier_short_effective}). The correction due to the second effect arises only in
the first iteration, $\theta^{(1)}$, and can be estimated as $\frac{eV}{\mu_0}
\frac{d}{R_c}\frac{d}{L}$. This turns out to be smaller than the correction
arising from change of the energy integration limits $\epsilon_\pm$ in
Eq.~(\ref{eq:current barrier_short_effective}) by a factor of $d/L$. Therefore,
in order to find the coefficient $\alpha_{cl}$ of the $V^2H$ term in the
nonlinear current to the leading order in $d/L$ we may replace the argument of
the sines in Eq.~(\ref{eq:current
  barrier_short_effective}) by their $H=0$ values. Doing so we obtain
in the leading order in $d/L$
\begin{equation}\label{eq:alpha_cl_result}
    \alpha_{cl}=\frac{3}{8\pi^2 \sqrt{2}} \frac{e^4 d^2}{\mu_0^2 mc}
\end{equation}
in agreement with the estimate Eq.~(\ref{eq:alpha_cl_estimate}).

While the ``rigid-wall'' confinement model applies to some
setups~\cite{Ensslin}, the use of gates and scanning probes results in
a ``soft'' confining potential changing over length scales of the
order of the size of the contact~\cite{Westervelt}. This difference
affects only the numerical coefficient in $\alpha_{cl}$.

In conclusion, we have shown that the nonlinear $I$-$V$ characteristic
of a classical ballistic system of interacting electrons lacks the
${\bf H}\to -{\bf H}$ symmetry. The classical contribution to the odd
in $H$ current, Eq.~(\ref{eq:alpha_cl_estimate}), stems from bending
of electron trajectories by a magnetic field, rather than from the
flux sensitivity of electron wave interference
pattern~\cite{Spivak2004,Sanches2004}. For a ballistic structure in
which electron motion is not chaotic the magnitudes of the classical
and interference contributions for weak fields $H$ and low
temperatures $T$ are of the same order. The characteristic scales of
$H$- and $T$- dependence for the classical contribution, however,
exceed significantly the corresponding scales for the interference
one.

This work was supported in part by NSF grants DMR 02-37296 and DMR
04-39026 (L.I.G.) and by the David and Lucille Packard Foundation
(A.V.A.). We are grateful to M. B\"{u}ttiker and B. Spivak  for
useful discussions. The hospitality of MPIPKS-Dresden and Aspen
Center for Physics is gratefully acknowledged.


\begin{thebibliography}{99}
\bibitem{footnote} The estimates of $\alpha_q$ and $\alpha_{cl}$
are valid if inelastic scattering is absent. This requires a
sufficiently long inelastic mean free path, $l_{in}(T) \gg d$,
yielding the restriction $T \lesssim \sqrt{\epsilon_F E_T}$ (here we
accounted for the dominant energy relaxation mechanism provided by
electron interactions).
\end{thebibliography}
\end{document}